# Non-proportional response between 0.1-100keV energy by means of highly monochromatic synchrotron X-rays

Ivan V. Khodyuk, Johan T. M. de Haas, and Pieter Dorenbos, *Member, IEEE*

*Abstract*— Using highly monochromatic X-ray synchrotron irradiation ranging from 9 keV to 100 keV, accurate $Lu_2SiO_5:Ce^{3+}$,Ca (LSO), $Lu_3Al_5O_{12}:Pr^{3+}$ (LuAG), $Lu_2Si_2O_7:Ce^{3+}$ (LPS) and $Gd_2SiO_5:Ce^{3+}$ (GSO) non-proportional response curves were determined. By utilizing information from escape peaks in pulse height spectra the response curve can be extended down to several keV. A detailed study of the non-proportionality just above the K-edge is presented and from that a method, which we named *K-dip spectroscopy*, is obtained to reconstruct the electron response curve down to energies as low as 100 eV.

*Index Terms*— Electron response, K-dip spectroscopy, non-proportionality, scintillator, X-ray response

## I. INTRODUCTION

IT was found 50 years ago [1] that the amount of light emitted in the scintillation spark caused by absorption of an X-ray or a γ-quantum in a crystal is not precisely proportional to its energy. This finding appears important because it causes the energy resolution achievable with scintillation material to deteriorate [2]. Although the phenomenon of non-proportional response and its relation with energy resolution has been studied quite intensively [3] - [9] there are still many major gaps in our understanding of the underlying physics. New theoretical models and accurate data by dedicated experimental techniques are needed to reveal the true origin of energy relaxation and dissipation inside the solid state. Better knowledge of the fundamental mechanisms of energy loss is necessary in the search for new highly effective and low energy resolution scintillators.

Non-proportional response as function of gamma energy is a direct consequence of the more fundamental non-proportional response to electrons. A powerful method to study the electron response of a scintillator is the Compton Coincidence Technique (CCT) introduced by Valentine and Rooney [10] and further developed by Choong *et al.* [11]. The main advantages of this method are the wide energy range covered, and the results are not affected by surface effects.

The authors are with Faculty of Applied Sciences, Delft University of Technology, Radiation Detection and Medical Imaging Group, Mekelweg 15, 2629 JB Delft, The Netherlands (e-mail: i.v.khodyuk@tudelft.nl; j.t.m.dehaas@tudelft.nl; p.dorenbos@tudelft.nl).

Nevertheless, using CCT, it is not possible to obtain accurate data on the electron response at energies below 3 keV. We will demonstrate in this work that measuring the photon response using highly monochromatic synchrotron X-rays, it is possible to get information of electron response starting from energy as low as 100 eV avoiding influence of surface effects. We are not aware of any other experimental method that provides information on electron response down to that low energy. Accurate experimental data is especially important in this low energy range because there the most dramatic drop in scintillator efficiency is expected.

In this work we will start from the non-proportionality response curves determined using direct observation of photo peaks from total absorption of highly monochromatic X-ray synchrotron irradiation; 9 – 100 keV X-rays were used. Typically 5 keV step size was used, much finer step size of 25 eV was used around the K absorption edge of the high Z atom in the scintillators. For each X-ray energy, energy resolution is determined as well. Next, a method to obtain the photon response curve in the low energy range down to 5 keV using $K_\alpha$ and $K_\beta$ escape peaks is presented. The non-proportionality curves as function of deposited energy are obtained for $Lu_2SiO_5:Ce^{3+}$,Ca (LSO), $Lu_3Al_5O_{12}:Pr^{3+}$ (LuAG), $Lu_2Si_2O_7:Ce^{3+}$ (LPS) and $Gd_2SiO_5:Ce^{3+}$ (GSO). Analysis of detailed data of the non-proportionality just above the K-edge, a method that we call *K-dip spectroscopy*, makes it possible to reconstruct the electron response curve that starts already at energies as low as 100 eV. Finally, a comparison of all three methods is presented for LSO and LuAG. The limitations of the methods and differences are discussed.

We would like to state that this work presents a first attempt to use the K-dip spectroscopy and escape analysis, in order to understand causes of non-proportional response in inorganic scintillators. The aim of this work is to provide new data and methods to obtain those. It is not our aim to provide a complete explanation of the observed non-proportional response curves.

## II. MATERIALS AND EXPERIMENT

### A. Scintillating materials

Table I compiles the studied samples. We decided to use these scintillators because LSO and GSO have similar chemical composition and crystal structure, and differ mainly by the atomic number of the lanthanide. LSO:Ce codoped with Ca is a scintillator with high density 7.4 g/cm$^3$ and light yield of 38800 photons per MeV of absorbed gamma ray energy (ph/MeV) [12]. In spite of high photon yield, the energy resolution at 662 keV of the best LSO sample is not better than 7%. This is attributed to a high degree of intrinsic non-proportionality of the material. GSO with four times lower photon yield of 10500 ph/MeV [13] has a comparable energy resolution of 8%. LPS was chosen because it like LSO contains Lu and has relatively high photon yield of 26000 ph/MeV, and at the same time poor energy resolution of 10% [14]. LuAG has been chosen because it is also a Lu-based compound and it is a promising new scintillator material with good proportionality and low energy resolution of 4.6% at 662 keV [15]. Relatively thin samples 1 – 2 mm thick were used, to increase the probability of $K_\alpha$ and $K_\beta$ X-ray escape events, which we are interested in.

TABLE I
THE CRYSTALS USED IN THE EXPERIMENT

| Crystal | Size, mm$^3$ | Resolution at 662 keV, % | Photon yield, photons/MeV | Reference |
|---|---|---|---|---|
| LuAG:Pr | $6 \times 6 \times 1$ | 4.6 | 19000 | [13] |
| GSO:Ce | $14 \times 10 \times 1.5$ | 8.1 | 10500 | [11] |
| LSO:Ce,Ca | $10 \times 10 \times 2$ | 7.7 | 38800 | [10] |
| LPS:Ce | $6 \times 7 \times 2$ | 10 | 26000 | [12] |

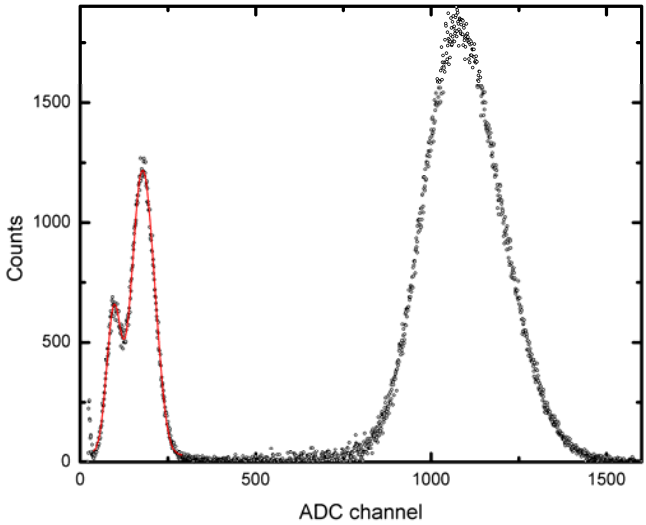

Fig. 1. Pulse height spectrum of LSO recorded using 70 keV monochromatic X-rays. The solid red line is the result of a fit of the escape peaks with five Gaussian peaks.

### B. Radioactive sources

Gamma ray pulse height spectra from radioactive sources were measured with standard spectroscopic techniques. The number of photoelectrons per MeV of absorbed gamma ray energy produced by the tested sample in a Hamamatsu R1791 photomultiplier tube was determined by comparing the peak position of the $^{137}$Cs 662 keV and $^{241}$Am 59.5 keV photopeaks in the pulse height spectra with the mean value of the single photoelectron peak position. The procedure has been described in detail by de Haas *et al.* [16]. For the optimization of the observed yield and energy resolution, the shaping time was set to 1μs.

### C. X-ray monochromator

To obtain more detailed non-proportionality curves, experiments at the X-1 beamline at the Hamburger Synhrotronstrahlungslabor (HASY – LAB) synchrotron radiation facility in Hamburg, Germany were carried out. A highly monochromatic pencil beam in the energy range 9 – 100 keV was used as excitation source. A tunable double Bragg reflection monochromator using a Si[511] and Si[311] set of silicon crystals providing an X-ray resolution of 1 eV at 9 keV rising to 20 eV at 100 keV was used to select the X-ray energies. The beam spot size was set by a pair of precision stepper-driven slits, positioned immediately in front of the sample coupled to the photomultiplier tube (PMT). For all measurements, a slit size of 50 × 50 μm$^2$ was used. The PMT was mounted on an X-Y table capable of positioning with a precision of <1 μm in each direction. Prior to each measurement session, the position of the PMT was adjusted to achieve as high count rate as possible. The intensity of the synchrotron beam was reduced in order to avoid pulse pileup. A lead shielding was used to protect the sample from background irradiation.

Figure 1 shows 70 keV synchrotron X-ray pulse height spectrum of LSO recorded with a Hamamatsu R1791 PMT connected to a homemade preamplifier, an Ortec 672 spectroscopic amplifier and an Amptek 8000A multichannel analyzer (MCA). Sample was optically coupled to the window of the PMT with Viscasil 600 000 cSt from General Electric. To improve the collection of scintillation light, the crystal was covered with several layers of ultraviolet reflecting Teflon tape (PFTE tape) forming an "umbrella" configuration [17]. All measurements were carried out at room temperature.

Corrections were made for channel offsets in the pulse height measurement. The offset was measured by the use of an Ortec 419 precision pulse generator with variable pulse height attenuation settings.

## III. RESULTS AND DISCUSSION

### A. Non-proportionality

The photon response of the scintillators was determined using monoenergetic X-rays with energies between 9 and 100 keV with a 5 keV step size. In the low energy range 9 to 15

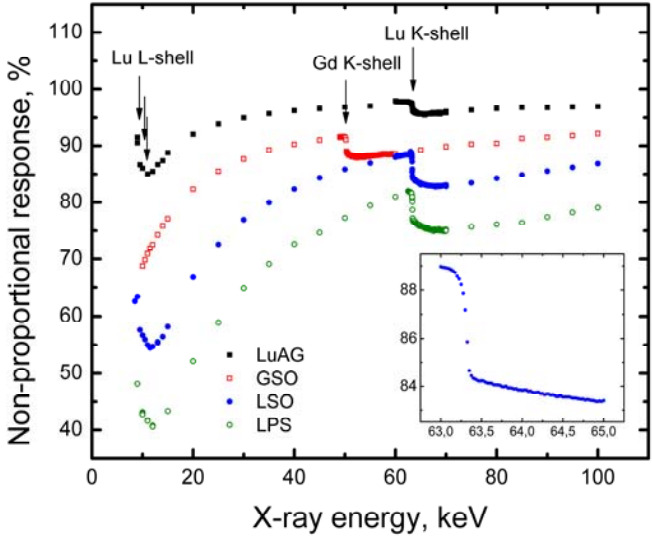

Fig. 2. Scintillation photon yield at RT as a function of X-ray energy for LuAG, GSO, LSO and LPS, relative to the photon yield at 662 keV excitation. The inset shows a 25 eV step size energy scan between 63 and 65 keV for LSO. Error bars are not shown, because the size of the error bar is comparable with the symbol size

keV, a 1 keV step size was used. A much finer step size of 25 eV was used near the K-shell electron binding energy of 63.314 keV for Lu or 50.239 keV for Gd-based compounds. Figure 2 shows the relative scintillation photon yield as a function of X-ray energy for LuAG, GSO, LSO and LPS. The nonproportional response is defined as the photoelectron yield/MeV at energy E divided by the photoelectron yield/MeV at 662 keV and presented in percents. Error bars are not shown in Fig. 2, because the size of the error bars is comparable with the symbols size. Precision of the experimental data can be seen in the inset of Fig. 2 and in case of LSO δ is less then 0.05% in the entire energy range.

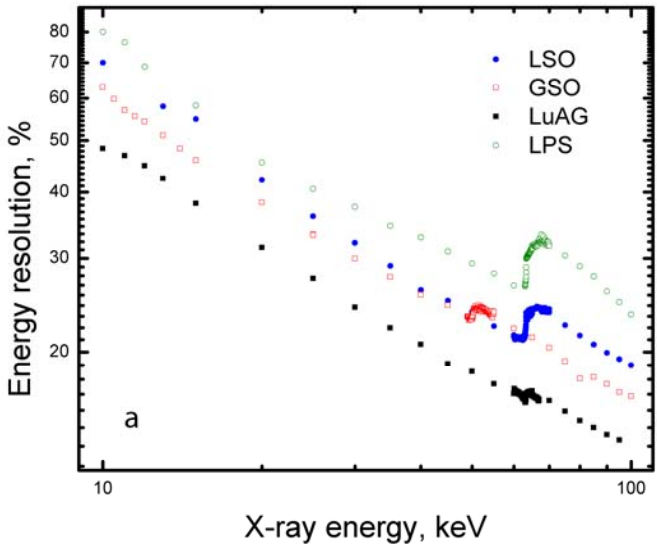

Fig. 3. Energy resolution versus X-ray energy. Error bars are not shown, because the size of the error bar is comparable with the symbol size.

TABLE II
ELECTRON BINDING ENERGIES

| Element | Electron binding energies, keV | | | |
|---|---|---|---|---|
| | K, 1s | $L_I$, 2s | $L_{II}$, $2p_{1/2}$ | $L_{III}$, $2p_{3/2}$ |
| Lutetium | **63.314** | 10.870 | 10.349 | 9.244 |
| Gadolinium | **50.238** | 8.376 | 7.930 | 7.243 |

All four materials reveal similar features. When moving from high energy towards low energy, we observe a relatively slow decrease of proportionality down to the K-edge energy. Table II compiles the K - and L -shell electron binding energies for Lu and Gd. Figure 2 shows that at the Lu or Gd K-edge energy the non-proportionality curve increases with a clear discontinuity at the edge. For LuAG the response at the edge increases by about 2.1%, while for GSO it is 3.6%. LSO has a change in response of 5.9% and this increases to 6.8% for LPS. Moving further towards lower energy we observe a drop of proportionality below the K-edge. At 9 keV the efficiency has decreased by 15%, 31%, 45%, and 60% for LuAG, GSO, LSO, and LPS, respectively. For the Lu-based compounds there are also discontinuities in the non-proportionality curve at the Lu L-shell energy.

We observe a proportional dependence between the magnitude of the drop $KL_{drop}$ of scintillator efficiency from below the K-shell to above the L-shell energy with the magnitude $K_{dip}$ of the drop at the K-edge. Empirically we can write

$$KL_{drop} = \xi \times K_{dip}, \quad (1)$$

where for the Lu-based compounds $\xi = 6.01 \pm 0.04$.

*B. Energy resolution*

The energy resolution, defined as Full Width at Half Maximum intensity (FWHM) over the peak position, determined from photopeaks like in Fig. 1, are plotted in Fig. 3 versus the X-ray energy and in Fig. 4 versus the number of photoelectrons. The solid curve in Fig. 4 represents the theoretical limiting resolution due to Poisson statistics in the number of detected photons [18]:

$$R_{stat} = 2.355 \sqrt{\frac{1 + \nu(M)}{N_{phe}}}, \quad (2)$$

where $\nu(M) = 0.15$ is the variance in the PMT gain.

Fig. 4 shows that the energy resolution for GSO and LuAG are quite close to the theoretical limit. For GSO the most important factor limiting energy resolution is the relatively low photon yield of 10500 photons/MeV. For LuAG the contribution from non-proportional response is very small. For LSO and LPS the values of their energy resolution are much further from the theoretical limit which is attributed to a large contribution from the non-proportional response.

An *S* type structure can be observed in the data at the Lu or Gd K-edges, see inset Fig. 4. For the LSO sample light yield starts to decrease approximately 300 eV below the Lu K-edge. With further increase of the X-ray excitation energy, the number of photoelectrons falls rapidly from 662 at 63.0 keV to 631 at 63.5 keV. A discontinuous increase of the energy resolution appeared in the same energy range 63.0 – 63.5 keV. Analogous features were observed for the other samples. LuAG shows the smallest resolution jump of only 0.9% at the K shell binding energy. For GSO the resolution jump is 1.4%, for LSO 3.2%, and for LPS 5.6%. There is not a linear relation between the magnitude of the resolution jump at the K-edge and $K_{dip}$, but a correlation clearly exists; the larger the resolution jump the larger the value for $K_{dip}$.

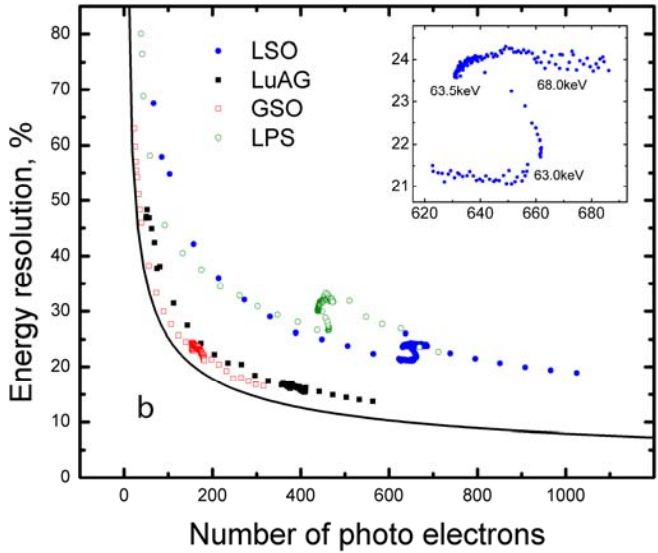

Fig. 4. Energy resolution versus number of photo electrons. The inset shows the *S* type structure around the Lu K-shell in the case of LSO.

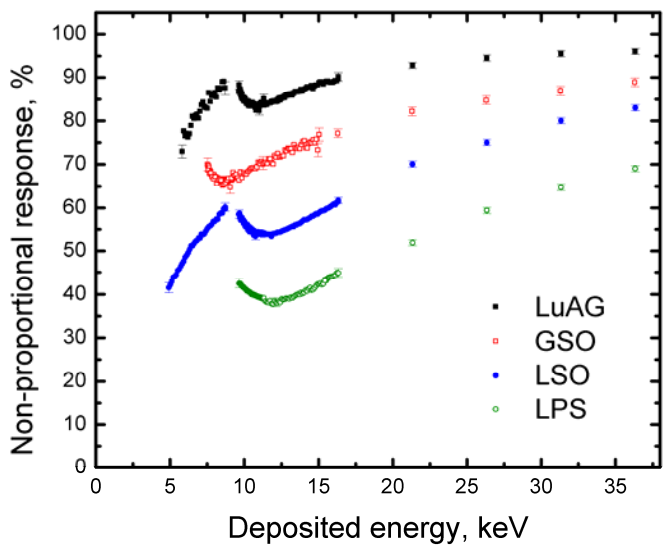

Fig. 5. Scintillation photon yield relative to the photon yield at 662 keV. reconstructed using escape peaks as a function of deposited energy for LuAG, GSO, LSO and LPS.

## C. Information derived from escape peaks

X-ray photons of energy between 9 keV and 100 keV interact with a sample almost exclusively by means of the photoelectric effect with a K-shell or L-shell electron. The electron is ejected from its shell, leaving a hole. As the atom returns to its stable lowest energy state, electrons from the outer shells relax to the inner shells, and in the process giving off a characteristic X-ray or Auger electrons. In the case that a characteristic X-ray photon escapes the bulk of the crystal we observe a so-called escape peak, shown in Fig. 1. The energy $E_{deposited}$ deposited in the bulk of the material is then:

$$E_{deposited} = E_X - E_{escape}, \qquad (3)$$

where $E_X$ is the energy of the incident X-ray photon and $E_{escape}$ - the energy of the fluorescent X-ray that escaped from the bulk of the material [19]. The ranges of Auger electrons are too short to escape the bulk of the material and we do not consider Auger electron escape here.

Using (3), we reconstructed the response curves for LuAG, GSO, LSO and LPS scintillators as shown in Fig. 5. Procedure described by us in detail in [20]. The parts of the curves at energies above the L-edge are obtained by analyzing $K_{\alpha 1}$ and $K_{\alpha 2}$ – escape peaks. $K_\alpha$ X-ray fluorescence is caused by a transition of an electron from one of the subshells of the L-shell to the hole in the K-shell. In the case of transition from the M or N-shell to the hole in the K-shell, a $K_{\beta 1}$, $K_{\beta 2}$ or $K_{\beta 3}$ X-ray can be emitted and by utilizing the $K_\beta$ escape peaks the photon response below the L-edge can be reconstructed for LuAG and LSO. The probabilities of other transitions leading to X-ray escape are very small and not considered here. It was not possible to resolve $K_\alpha$ and $K_\beta$ escape peaks for GSO and LPS, and hence no data below the L-edges could be retrieved. In the case of LPS the absence of the $K_\beta$ escape peak is due to a very poor energy resolution, and for GSO because of the lower energy of the Gd L-edge compared to that of Lu. The statistical contribution to the energy resolution becomes large in the low X-ray energy range and a high photon yield of the material is then important to be able to resolve the $K_\alpha$ and $K_\beta$ escape peaks.

Reconstructed data match data obtained from photopeak analysis well like in [21]. Rapid changes of the scintillator efficiency are observed near energies of Lu and Gd L-shell. A sharp discontinuity like in the case of the K-edge is not seen because there are three different L-subshell electron binding energies instead of only one for the K-shell.

## D. K-dip spectroscopy

Analysis of data on the non-proportionality response just above the K-edge Fig. 6, makes it possible to reconstruct the electron response curve that starts already at energies as low as 100 eV. We call this method *K-dip spectroscopy*. Briefly, the method can be described as follows. The response of a scintillator to an X-ray that has interacted with a K-shell electron is equivalent to the response of a scintillator to a sum of two main interaction products: 1) a K-shell photo electron response plus 2) the sequence of processes following relaxation of the hole in the K-shell, the so-called K-cascade response. Our strategy is to employ X-ray energies just above the K-edge. The scintillator response due to the K-cascade is assumed independent from the original X-ray energy. This response is found by tuning the X-ray energy very close above the K-shell binding energy. By subtracting the K-cascade response from the total X-ray response we are left with the response in photoelectrons/MeV from the K-shell photo-electron alone with energy $E_X - E_{K_{binding}}$. To estimate the scintillator response due to the K-cascade, we fitted experimental data just above the K electron binding energy with polynomial fit Fig. 6. The mean value of the fit function at 63.314 keV was taken as light yield produced by K-cascade. The non-proportionality curve is then obtained by dividing with the photoelectron yield/MeV at energy of 662 keV. This procedure is analogous to the one developed by Collinson and Hill [22].

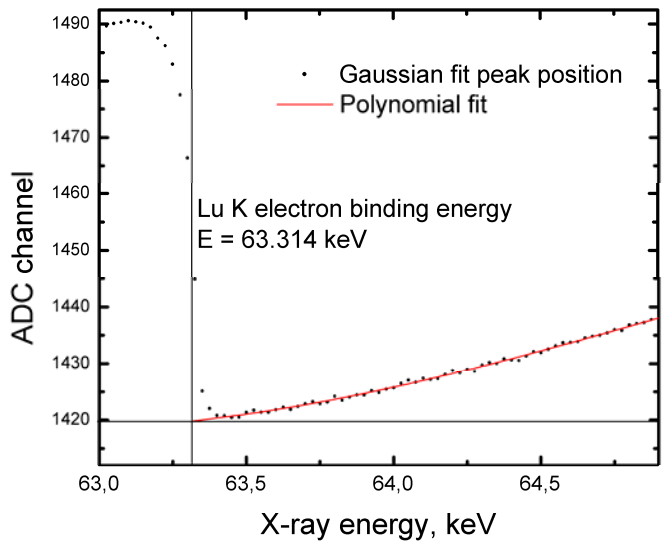

Fig. 6. Position of the Gaussian fitted photopeak for LSO as a function of X-ray energy. Solid red line represents polynomial fit of the data above the Lu K electron binding energy.

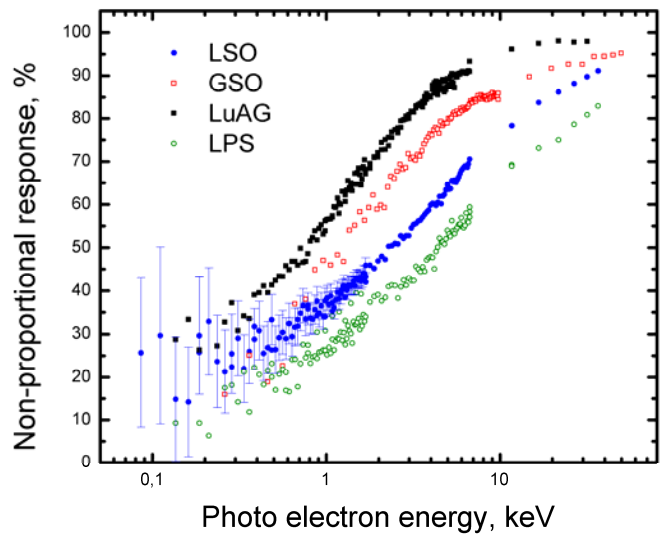

Fig. 7. Relative photon yield as a function of photo electron energy inferred using K-dip spectroscopy.

Figure 7 shows the results obtained from this K-dip spectroscopy method. Our method is good enough to show shape and principal behavior of the electron response curve. The error bars, shown in Fig. 7 for LSO are substantial because of subtracting of two close quantities. We are not presenting error bars for the other samples in order to keep data readability.

A drop of 70% to 90% in the scintillator efficiency can be seen for the samples. Like in Fig. 2 we have the same ordering of the curves in Fig. 7; the most proportional one is for LuAG, then for GSO, LSO, and finally for LPS. Fig. 7 reveals a new finding. For all Lu based samples the slope of the electron non-proportional response curve tends to become less steep below 1 keV.

According to theory [23, 24] on electron inelastic mean free paths and stopping powers, the deposited energy density $dE/dx$ along the track as a function of electron energy increases rapidly with decrease of electron energy. In the low energy range peaking of the deposited energy density is observed in condensed matter. Position of the density maximum is material dependent, but is in order of few hundreds of electron volts. Since a high deposited energy density $dE/dx$ is thought to be the main reason for scintillator efficiency losses [2], [3], we expect a relationship between the non-proportionality curve and the deposited energy density curve.

*E. Comparison of the three methods*

Fig. 8 and Fig. 9 compare the three sets of response curve data of LSO and LuAG, obtained from X-ray photopeak position analysis, escape peak analysis, and K-dip spectroscopy. As shown in the insets, the shapes of the curves obtained from the photopeak and escape peaks agree well but do have small differences. Those differences we attribute to the fact that the atomic state after photoelectric interaction with the L-shell is different from the atomic state after X-ray fluorescence. In the former situation a hole is created with equal probability in the $L_I$, $L_{II}$, and $L_{III}$ subshells. In the latter situation the hole is preferably created in the $L_{III}$ and $L_{II}$ subshells [19]. The cascade products and the resulting photon yields are then not necessarily equal for both situation leading to the small differences in the data from both methods in Fig. 6 and 7. In the entire energy range, the K-shell photo-electron response curves are located above the X-ray photon response curves. It means that a single electron produces higher number of scintillating photons then an X-ray or gamma photon of the same energy. It can be understood by comparing the secondary reaction products from single electron and photon interaction in the sample. An X-ray photon, depending on its energy, can be photoelectrically absorbed by one of the atom's shells, creating a relatively energetic photoelectron and a set of electrons of low energies as a result of the cascade process following hole relaxation. Because of a high non-proportionality of the scintillator in the low energy region, the number of photons created by the set of low energy electrons will always be smaller then the yield from the single electron of the same total energy.

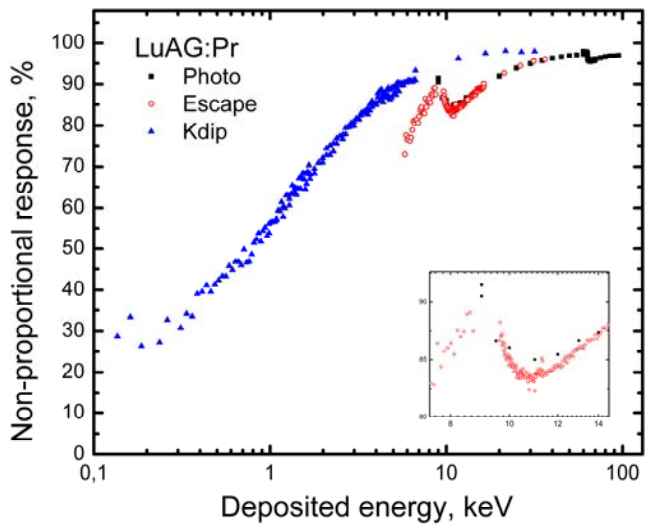

Fig. 8. Comparison of direct photon response, response reconstructed using escape peaks and electron response inferred using K-dip spectroscopy for LuAG. The inset shows zoomed in part of the curves near L binding energies.

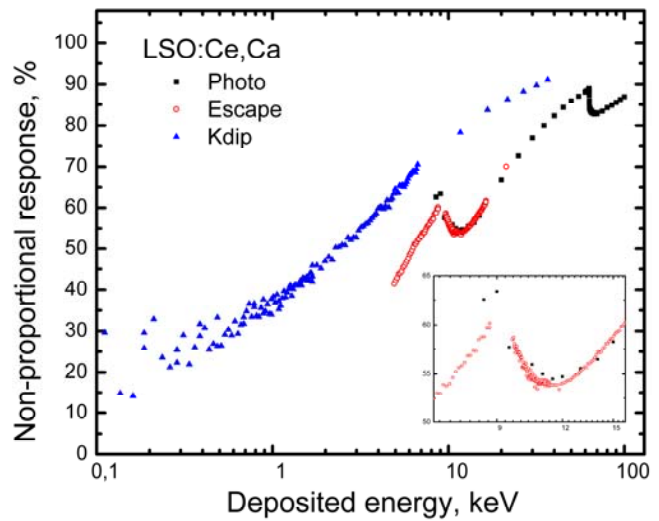

Fig. 9. Comparison of direct photon response, response reconstructed using escape peaks and electron response inferred using K-dip spectroscopy for LSO. The inset shows zoomed in part of the curves near L binding energies.

## IV. CONCLUSION

We have measured the scintillation response of LuAG, GSO, LSO, and LPS to X-rays in the energy range 9 – 100 keV, with special emphasis near the K-edge of the high Z atoms in the scintillator. From this data, we have inferred the electron response curves of the materials in the energy range 0.1 – 30 keV by a new method that we named K-dip spectroscopy. From 30 keV to 1 keV, scintillation yield for all samples appears to drop by 50 to 75%. Below 500 eV the response becomes proportional again. A means to construct the low energy photon response using analysis of escape peaks has been presented.

Our methods utilizing escape peaks and K-dip spectroscopy have the advantage that the non-proportionality curve can be extended to lower energies than possible with other methods. CCT becomes too inaccurate below 3 keV. With K-dip spectroscopy the curves are extended down to 100 eV. Detailed study of the non-proportionality in the photon response just above the K-edge using energy step as small as 25 eV makes this possible.

The CCT method has an advantage over K-dip spectroscopy. In K-dip spectroscopy we suppose that in the K-cascade a set of low energy electrons are emitted from the atom and each produces an ionization track. We assumed that these tracks do not interfere with the track created by the K-shell photoelectron. In that case the K-dip spectroscopy method provides us like the CCT method the genuine electron response. However, when tracks do influence each other, i.e. when the number of photons produced by the photoelectron is affected by the tracks from the cascade products, an error is introduced In this regard CCT may have an intrinsic advantage over the K-dip spectroscopy, by exciting the crystal with essentially just one electron at a time.

### ACKNOWLEDGMENT

Authors would like to thank Alan Owens and Francesco Quarati for providing beamtime and assisting with experiment at the X-1 beamline at the Hamburger Synhrotronstrahlungslabor (HASY – LAB) synchrotron radiation facility. These investigations have been supported by The Netherlands Technology Foundation (STW)